# Phonons become chiral in the pseudogap phase of cuprates

G. Grissonnanche[1] *, S. Thériault[1], A. Gourgout[1], M.-E. Boulanger[1], E. Lefrançois[1],
A. Ataei[1], F. Laliberté[1], M. Dion[1], J.-S. Zhou[2], S. Pyon[3,4], T. Takayama[3,5],
H. Takagi[3,5,6,7], N. Doiron-Leyraud[1], and L. Taillefer[1,8] *

*1 Département de physique, Institut quantique, and RQMP, Université de Sherbrooke, Sherbrooke, Québec, Canada*

*2 Materials Science and Engineering Program, Department of Mechanical Engineering, University of Texas at Austin, Austin, TX, USA*

*3 Department of Advanced Materials Science, University of Tokyo, Kashiwa, Japan*

*4 Department of Applied Physics, University of Tokyo, Tokyo, Japan*

*5 Max Planck Institute for Solid State Research, Stuttgart, Germany*

*6 Department of Physics, University of Tokyo, Tokyo, Japan*

*7 Institute for Functional Matter and Quantum Technologies, University of Stuttgart, Stuttgart, Germany*

*8 Canadian Institute for Advanced Research, Toronto, Ontario, Canada*

**The nature of the pseudogap phase of cuprates remains a major puzzle[1,2]. One of its new signatures is a large negative thermal Hall conductivity $\kappa_{xy}$, which appears for dopings $p$ below the pseudogap critical doping $p^*$, but whose origin is as yet unknown[3]. Because this large $\kappa_{xy}$ is observed even in the undoped Mott insulator $La_2CuO_4$, it cannot come from charge carriers, these being localized at $p = 0$. Here we show that the thermal Hall conductivity of $La_2CuO_4$ is roughly isotropic, being nearly the same for heat transport parallel and normal to the $CuO_2$ planes, i.e. $\kappa_{zy}(T) \approx \kappa_{xy}(T)$. This shows that the Hall response must come from phonons, these being the only heat carriers able to move as easily normal and parallel to the planes[4]. At $p > p^*$, in both $La_{1.6-x}Nd_{0.4}Sr_xCuO_4$ and $La_{1.8-x}Eu_{0.2}Sr_xCuO_4$ with**

**$p = 0.24$, we observe no *c*-axis Hall signal, *i.e.* $\kappa_{zy}(T) = 0$, showing that phonons have zero Hall response outside the pseudogap phase. The phonon Hall response appears immediately below $p^* = 0.23$, as confirmed by the large $\kappa_{zy}$ signal we find in $La_{1.6-x}Nd_{0.4}Sr_xCuO_4$ with $p = 0.21$. The microscopic mechanism by which phonons become chiral in cuprates remains to be identified. This mechanism must be intrinsic – from a coupling of phonons to their electronic environment – rather than extrinsic, from structural defects or impurities, as these are the same on both sides of *p\**. This intrinsic phonon Hall effect provides a new window on quantum materials[5] and it may explain the thermal Hall signal observed in other topologically nontrivial insulators[6,7].**

The thermal Hall effect has emerged as a new probe of insulators[8,9], materials in which the electrical Hall effect is zero because there are no mobile charge carriers. In the presence of a heat current $J$ along the $x$ axis and a magnetic field $H$ along the $z$ axis, a transverse temperature gradient $\nabla T$ (along the $y$ axis) can develop even if the carriers of heat are neutral (chargeless), provided they have chirality[10]. In particular, carriers with a Berry curvature – whether fermions or bosons – will in general generate a nonzero thermal Hall conductivity $\kappa_{xy}$ (refs. 11,12). So, potentially, a measurement of the thermal Hall effect can provide access to various topological excitations in quantum materials, for example Majorana edge modes in chiral spin liquids[13,14,15].

However, it turns out that phonons can also generate a nonzero thermal Hall conductivity, if some mechanism – intrinsic[11] or extrinsic[16] – confers chirality to the phonons. (Here we use the term "chirality" to mean handedness in the presence of a magnetic field.) The phonon $\kappa_{xy}$ signal can be large, as in multiferroic materials[17] – where the mechanism is spin-phonon coupling – or in strontium titanate ($SrTiO_3$) (ref. 18) – where the mechanism seems to involve structural domain boundaries.

In cuprates, a large negative $\kappa_{xy}$ signal was observed at low temperature inside the pseudogap phase, *i.e.* for dopings $p < p^*$ (ref. 3). Because it persists down to $p = 0$, in the Mott insulator state, this negative $\kappa_{xy}$ cannot come from charge carriers. Therefore, it must come either from spin-related excitations (possibly topological, as in ref. 19) or from phonons. To distinguish between these two types of heat carriers, we adopt a simple approach: we measure the thermal Hall conductivity for a heat current along the *c* axis, normal to the $CuO_2$ planes, a direction in which only phonons move easily[4] (see Methods).

We start by looking at the undoped cuprate $La_2CuO_4$, a Mott insulator with no mobile charge carriers. Here, phonons are the dominant heat carriers at low *T* and their longitudinal thermal conductivity $\kappa_{nn}(T)$ is nearly the same for *J* // *a* ($n = x$) and *J* // *c* ($n = z$) (ref. 4; see Methods). As reproduced in Fig. 1a, the in-plane thermal Hall conductivity $\kappa_{xy}$ of $La_2CuO_4$ (*J* // -*x* ⊥ *c*, *H* // *z* and ∇*T* // *y*; Extended Data Fig. 1a) was previously found to be negative at all *T*, with $|\kappa_{xy} / T|$ growing steadily as temperature is reduced below 100 K, reaching one of the largest Hall conductivities of any insulator, at *T* = 10 K (ref. 3). In a separate sample of $La_2CuO_4$, we measured $\kappa_{zy}$ (*J* // *z* // *c*, *H* // *x* and ∇*T* // *y*; Extended Data Fig. 1b) and found that $\kappa_{zy}(T) \approx \kappa_{xy}(T)$ at all *T* (Fig. 1a). The fact that the thermal Hall conductivity is as large across the $CuO_2$ planes as within is compelling evidence that the carriers of heat responsible for the thermal Hall effect in $La_2CuO_4$ are phonons. Indeed, any excitation of electronic origin (carrying charge or spin) is expected to be much more mobile within the $CuO_2$ planes as opposed to across planes (see Methods).

Next, we turn to the hole-doped cuprate $La_{1.6-x}Nd_{0.4}Sr_xCuO_4$ (Nd-LSCO), whose phase diagram is shown in Fig. 2a. The pseudogap phase boundary $T^*(p)$ ends at a $T = 0$ critical point $p^* = 0.23$, as determined by both transport[20] and photoemission (ARPES; ref. 21) measurements. At $p = 0.24$, just above $p^*$, the in-plane thermal Hall conductivity $\kappa_{xy}$ of Nd-LSCO was found to be positive at all $T$ (Fig. 1b) and in good agreement with the Wiedemann-Franz law, namely $\kappa_{xy} / T = L_0 \sigma_{xy}$ as $T \rightarrow 0$, where $L_0 = (\pi^2/3)(k_B / e)^2$ (ref. 3). Using a sample cut from the same large crystal, we measured $\kappa_{zy}$ and found that $\kappa_{zy}(T) = 0$ at all $T$, below 100 K (Fig. 1b). Combined with the fact that the Wiedemann-Franz law is satisfied for $J // a$, our data for $J // c$ show that phonons in Nd-LSCO have no Hall effect at $p = 0.24$, for any direction. (Note that the contribution of charge carriers to $\kappa_{zy}(T)$ is extremely small – see Methods.) In other words, phonons have no chirality outside the pseudogap phase.

In the related material $La_{1.8-x}Eu_{0.2}Sr_xCuO_4$ (Eu-LSCO), also with $p^* = 0.23$ (ref. 22), we again find that $\kappa_{zy}(T) = 0$ at $p = 0.24 > p^*$ (Fig. 1d). We stress that our data showing $\kappa_{zy}(T) = 0$ down to 10 K (on two separate samples) demonstrate that our measurement technique does not introduce any spurious background Hall signal (see Methods). In other words, any thermal Hall signal coming from the sample mount is negligible compared to the signal due to the samples.

Inside the pseudogap phase ($p < p^*$), a large negative $\kappa_{xy}$ was observed at low $T$ in $La_{2-x}Sr_xCuO_4$ (LSCO) with $p = 0.06$, Eu-LSCO with $p = 0.08$, $Bi_2Sr_{2-x}La_xCuO_{6+\delta}$ with $x = 0.2$, and Nd-LSCO with $p = 0.20, 0.21$ and $0.22$ (ref. 3). In Fig. 1c, we reproduce the published $\kappa_{xy}$ data for Nd-LSCO with $p = 0.21$, seen to go negative below 25 K – in contrast to $\sigma_{xy}$, which remains positive down to $T \rightarrow 0$ (refs. 3,20). In the same figure, we report our data for $\kappa_{zy}$ measured in Nd-LSCO with $p = 0.21$. We see that in striking contrast with $p = 0.24$, we now have a sizable (negative) $\kappa_{zy}$ signal.

We summarize our $\kappa_{zy}$ measurements in Fig. 2b. At $p = 0.24$, just outside the pseudogap phase ($p > p^*$; Fig. 2a), $\kappa_{zy}(T) = 0$ and phonons have no chirality. At $p = 0.21$, just inside the pseudogap phase ($p < p^*$; Fig. 2a), $\kappa_{zy}(T) << 0$ and phonons have suddenly acquired chirality. This new phonon Hall effect grows in strength with decreasing doping, being largest at $p = 0$, in $La_2CuO_4$. We therefore have two key findings: the large negative thermal Hall signal in cuprates is carried by phonons and the phonons become chiral only once they enter the pseudogap phase. (In ref. 3, a phonon scenario was considered unlikely because of the smallness of two expected signatures: a field dependence of $\kappa_{xx}$ and a drop in $\kappa_{xx}$ below $p^*$. These quantitative considerations now have to be understood. See Methods.)

The question then becomes this: what special property of the pseudogap phase confers chirality to phonons? One possibility is that phonons acquire Berry curvature[11] from their interaction with the special electronic properties of that phase. A rather universal consequence of Berry curvature is to produce a thermal Hall response that varies as $\kappa_{xy} / T \sim \exp(-T/T_0)$ at intermediate temperatures (ref. 12). In Fig. 3, we show a fit of our $\kappa_{zy}$ data to the relation $\kappa_{zy} / T = A \exp(-T/T_0) + C$, for $La_2CuO_4$ and Nd-LSCO with $p = 0.21$. We see that the fits are excellent, down to $T \sim T_0 \sim 15$ K. This supports the scenario of phonons with Berry curvature (below $p^*$).

Further experimental and theoretical work is needed to identify the microscopic mechanism responsible for the chirality of phonons in the pseudogap phase. Note that it cannot simply be the skew scattering of phonons from impurities. Indeed, while skew scattering of phonons by magnetic impurities can produce a thermal Hall effect[16,23], typically very small (orders of magnitude smaller than that found in $La_2CuO_4$), this extrinsic impurity-related mechanism cannot apply here since for the same Nd-LSCO material (with the same impurities) we find zero thermal Hall effect when $p > p^*$. Also, changing non-magnetic Eu ions for magnetic Nd ions in $La_{2-y-x}RE_ySr_xCuO_4$ (RE=Eu,Nd), at $p$ = 0.24, still yields zero phonon Hall signal. What is needed is a qualitative change below $p^*$ in the intrinsic coupling of phonons to their environment. Interestingly, a recent ARPES study in the cuprate $Bi_2Sr_2CaCu_2O_{8+\delta}$ saw a rapid increase in the coupling of phonons to electrons upon crossing below $p^*$ (ref. 24).

A large $\kappa_{xy}$ signal due to phonons was recently observed in $SrTiO_3$, but not in the related material $KTaO_3$, where $\kappa_{xy}(T) \approx 0$ below 100 K (ref. 18). This striking difference was attributed to the presence of structural domains in $SrTiO_3$, absent in $KTaO_3$. Exactly how structural domains can generate a Hall effect is still unclear[25], but this mechanism cannot be responsible for the phonon Hall effect in Nd-LSCO. Indeed, there is no change in the crystal structure of Nd-LSCO between $p$ = 0.21 and $p$ = 0.24 – both are in the so-called LTT phase (see Methods).

A large $\kappa_{xy}$ signal due to phonons was observed in multiferroic materials like $Fe_2Mo_3O_8$ (ref. 17), where it was attributed to a coupling of phonons to spins. A spin-phonon coupling could be relevant in the case of cuprates, given that the pseudogap phase is characterized by short-range antiferromagnetic correlations and spin singlet formation, according to numerical solutions of the Hubbard model[26]. It may be that the topological character of this unusual state of correlated spins[27] confers chirality to phonons. Note that slow antiferromagnetic correlations (quasi-static moments) are indeed observed for dopings up to $p^*$ in Nd-LSCO (ref. 28) and LSCO (ref. 29), and not above.

The broad implication of our finding that phonons in insulators can generate large thermal Hall signals is to impose a re-examination of previous studies where the thermal Hall effect was attributed to heat carriers other than electrons or phonons, for example to Majorana edge modes in the 2D insulator α-$RuCl_3$ (ref. 30).

**Acknowledgements.** L.T. acknowledges support from the Canadian Institute for Advanced Research (CIFAR) as a Fellow and funding from the Natural Sciences and Engineering Research Council of Canada (NSERC; PIN: 123817), the Fonds de recherche du Québec - Nature et Technologies (FRQNT), the Canada Foundation for Innovation (CFI), and a Canada Research Chair. This research was undertaken thanks in part to funding from the Canada First Research Excellence Fund. Part of this work was funded by the Gordon and Betty Moore Foundation's EPiQS Initiative (Grant GBMF5306 to L.T.). J.-S.Z. was supported by an NSF grant (MRSEC DMR-1720595).

**Author contributions.** G.G., S.T., M.E.B., and E.L. performed the thermal Hall conductivity measurements. A.G., A.A., F.L., M.D. and N.D.-L. prepared and characterized the samples. J.-S.Z. grew the Nd-LSCO single crystals. S.P., T.T., and H.T. grew the Eu-LSCO and $La_2CuO_4$ single crystals. G.G. and L.T. wrote the manuscript, in consultation with all authors. L.T. supervised the project.

**Author information.** The authors declare no competing financial interest. Correspondence and requests for materials should be addressed to G.G. (gael.grissonnance@usherbrooke.ca) or L.T. (louis.taillefer@usherbrooke.ca).

## MAIN FIGURE CAPTIONS

**Fig. 1 | Thermal Hall conductivity of cuprates at three different dopings.**

Thermal Hall conductivity versus temperature in a field of magnitude $H$ = 15 T, plotted as $\kappa_{ny} / T$ vs $T$ for two heat current directions, $J // a$ ($n = x$; blue) and $J // c$ ($n = z$; red) in **a)** $La_2CuO_4$ ($p$ = 0), **b)** Nd-LSCO with $p$ = 0.24, **c)** Nd-LSCO with $p$ = 0.21, and **d)** Eu-LSCO with $p$ = 0.21. All lines are a guide to the eye. The $\kappa_{xy}$ data are from ref. 3; in Nd-LSCO, they were taken in 18 T, and so are multiplied here by a factor 15/18.

**Fig. 2 | Evolution of the *c*-axis thermal Hall conductivity across the phase diagram.**

**a)** Temperature-doping phase diagram of Nd-LSCO, showing the superconducting transition temperature $T_c$ (black line; zero field) and the pseudogap phase below $T^*$ (PG; orange region), which ends at the critical doping $p^* = 0.23$ (diamond) for both Nd-LSCO (refs. 2, 20) and Eu-LSCO (ref. 22). The orange circles indicate the temperature below which the in-plane resistivity deviates upwards from its *T*-linear dependence at high temperature (refs. 20, 31). The orange square marks the onset temperature for the opening of the anti-nodal pseudogap in Nd-LSCO at $p = 0.20$ detected by ARPES (ref. 21). The two vertical bands indicate the two dopings on either side of $p^*$ at which we measured $\kappa_{zy}(T)$, the *c*-axis thermal Hall conductivity shown in panel b (blue for $p = 0.21$, green for $p = 0.24$). **b)** Thermal Hall conductivity $\kappa_{zy}$, for a heat current normal to the $CuO_2$ planes ($J \mathbin{/\mkern-4mu/} c \mathbin{/\mkern-4mu/} z$) and a magnetic field of 15 T applied parallel to the planes ($H \mathbin{/\mkern-4mu/} a \mathbin{/\mkern-4mu/} x$), plotted as $\kappa_{zy} / T$ vs $T$, in $La_2CuO_4$ ($p = 0$; red), Nd-LSCO with $p = 0.21$ (blue), and Nd-LSCO with $p = 0.24$ (green).

**Fig. 3 | Phenomenological fit to the phonon thermal Hall conductivity.**

Fit of the thermal Hall conductivity $\kappa_{zy}(T)$ in **a)** $La_2CuO_4$ and **b)** Nd-LSCO at $p = 0.21$ to the phenomenological expression $\kappa_{zy} / T = A \exp(-T/T_0) + C$, derived from a theory that links the thermal Hall effect to the Berry curvature of the heat carriers[12]. The fit interval is from 15 K to 100 K. The resulting fit parameters are: a) $A = -4.9$ mW/K$^2$cm, $C = -0.02$ mW/K$^2$cm, $T_0 = 17.5$ K; b) $A = -2.4$ mW/K$^2$cm, $C = -0.04$ mW/K$^2$cm, $T_0 = 15.8$ K. The fact that the theoretical expression fits the data well in the temperature range above $T_0$ supports the hypothesis that phonons in those cuprates have a non-zero Berry curvature.

# METHODS

SAMPLES

**Nd-LSCO.** Single crystals of $La_{2-y-x}Nd_ySr_xCuO_4$ (Nd-LSCO) were grown at the University of Texas at Austin using a travelling-float-zone technique, with a Nd content $y = 0.4$ and nominal Sr concentrations $x = 0.21$ and 0.25. The hole concentration $p$ is given by $p = x$, with an error bar $\pm 0.003$, except for the $x = 0.25$ sample, for which the doping is $p = 0.24 \pm 0.005$ (for details, see ref. 20). The value of $T_c$, defined as the point of zero resistance, is: $T_c = 15$ and 11 K for samples with $p = 0.21$ and 0.24, respectively. The pseudogap critical point in Nd-LSCO is at $p^* = 0.23 \pm 0.005$ (ref. 20). The $a$-axis ($J // a$) and $c$-axis ($J // c$) samples were both cut out of the same large single crystal. The orientation and cutting of the samples were performed in Sherbrooke.

**Eu-LSCO**. The single crystal of $La_{2-y-x}Eu_ySr_xCuO_4$ (Eu-LSCO) was grown at the University of Tokyo using a travelling-float-zone technique, with a Eu content $y = 0.2$ and nominal Sr concentration $x = 0.24$. The hole concentration $p$ is given by $p = x$, with an error bar of $\pm 0.005$. The value of $T_c$, defined as the point of zero resistance, is $T_c = 9$ K. The pseudogap critical point in Eu-LSCO is at $p^* = 0.23 \pm 0.005$ (ref. 22). The $a$-axis ($J // a$) and $c$-axis ($J // c$) samples were both cut out of the same large single crystal. The orientation and cutting of the samples were performed in Sherbrooke.

**$La_2CuO_4$.** Our two single crystals of $La_2CuO_4$ came from the same batch, grown at the University of Tokyo using a travelling-float-zone technique. The $a$-axis ($J // a$) and $c$-axis ($J // c$) samples were each cut out of these two single crystals, respectively. The orientation and cutting of the samples were performed in Sherbrooke.

THERMAL HALL MEASUREMENT

Our measurements of the $c$-axis thermal Hall conductivity $\kappa_{zy}$ were performed on four samples: $La_2CuO_4$ ($p = 0$); Nd-LSCO with $p = 0.21$; Nd-LSCO with $p = 0.24$; Eu-LSCO with $p = 0.24$. In the same three materials, with the same four dopings, the in-plane thermal Hall conductivity $\kappa_{xy}$ was previously reported, in ref. 3. Those $\kappa_{xy}$ data are reproduced in the four panels of Fig. 1 (blue curves).

**Experimental Procedure.** For our measurements, six contacts were made on the sample using silver epoxy Dupont H20E diffused by annealing at high temperature in

oxygen – two contacts for the heat current, two for the longitudinal temperature difference $\Delta T_n$ ($n = x$ or $z$) and two for the transverse temperature difference $\Delta T_y$ (Extended Data Fig. 1). The sample was glued on a copper heat sink (Extended Data Fig. 1) with Dupont silver paint. Sample temperatures $T+$ and $T-$ are measured with one absolute type-E (chromel-constantan) thermocouple connected to $T-$ and one differential type-E thermocouple connected to $T+$ and $T-$ (which measures the temperature difference $\Delta T_n = T+ - T-$). Another differential type-E thermocouple measures the transverse temperature difference $\Delta T_y$ (Extended Data Fig. 1). A finite temperature difference $\Delta T_n$ is created by applying heat to the free end of the sample (Extended Data Fig. 1), using a 5 kΩ resistor whose resistance is well known and does not vary with temperature or field. Thermocouples and heater are connected to the sample with silver wires (25 and 50 μm in diameter, respectively).

The experiment is performed in a fixed magnetic field. A positive field $H = +15$ T is applied at $T = 100$ K and the sample is then cooled down to $T \sim 5$ K. At fixed field, the temperature is increased in steps, and the system is stabilized at each temperature point. For each point, the background voltages across all thermocouples in the absence of applied heat are carefully measured. Then heat is applied and, once the sample has reached thermal equilibrium, the thermocouple voltages are measured again. By subtracting the background voltages from the corresponding heat-on voltages, we can reliably extract the intrinsic response of the sample. We repeat this procedure at each temperature point from ~ 5 K to 100 K. Once the entire temperature range is covered at positive field, the field is inverted at $T = 100$ K, to its negative value $H = -15$ T, and the system is cooled down to ~ 5 K. We then repeat the whole procedure, under otherwise identical conditions. As always in a Hall measurement, the pure transverse signal (here $\Delta T_y$) is obtained by anti-symmetrization: $\Delta T_y = [\Delta T_y(+15\text{T}) - \Delta T_y(-15\text{T})] / 2$.

To determine the sign of the (transverse) thermal Hall conductivity, i.e. the sign of $\Delta T_y$, we use reference samples for which the sign of the thermal Hall conductivity is unambiguous, for example overdoped $Tl_2Ba_2CuO_{6+\delta}$ (positive sign), measured using the very same set-up (and thermocouple connections). For further details, see ref. 32.

For all the $\kappa_{zy}$ data reported here, the magnetic field strength was $H = 15$ T. Because $H \perp c$ in this case, superconductivity is only weakly suppressed, so in order to report only normal-state data we limit the data in Figs. 1, 2 and 3 to $T > T_c$. For the $\kappa_{xy}$ data reproduced in Fig. 1 (blue curves), the field strength was either $H = 15$ T (for $La_2CuO_4$ and Eu-LSCO) or $H = 18$ T (for Nd-LSCO) (ref. 3). In Figs. 1b and 1c, we have multiplied the $\kappa_{xy}$ data taken at 18 T by a factor 15/18, to enable a comparison at 15 T.

## ORIENTATION OF HEAT CURRENT AND MAGNETIC FIELD

**In-plane heat current ($J \perp c$).** For our prior measurements of $\kappa_{xy}$ (ref. 3), the heat current $J$ was sent in the basal plane of the single crystal (along -$x$), so parallel to the $CuO_2$ planes, generating a longitudinal temperature difference $\Delta T_x = T+ - T-$ (Extended Data Fig. 1a). The longitudinal thermal conductivity along the $x$ axis is given by $\kappa_{xx} = (J / \Delta T_x) (L_x / w_y t_z)$, where $L_x$ is the separation (along $x$) between the two points at which $T+$ and $T-$ are measured, $w_y$ is the width of the sample (along $y$) and $t_z$ its thickness (along $z \mathbin{/\mkern-5mu/} c$). By applying a magnetic field $H$ along the $c$ axis of the crystal (along $z$), normal to the $CuO_2$ planes, one generates a transverse temperature difference $\Delta T_y$ (Extended Data Fig. 1a). The in-plane thermal Hall conductivity is defined as $\kappa_{xy} = -\kappa_{yy} (\Delta T_y / \Delta T_x) (L_x / w_y)$, where $\kappa_{yy}$ is the longitudinal thermal conductivity along the $y$ axis. Here, we can take $\kappa_{yy} = \kappa_{xx}$, as all samples are either twinned or tetragonal.

**Out-of-plane heat current ($J \mathbin{/\mkern-5mu/} c$).** For the measurements of $\kappa_{zy}$ presented here, the heat current $J$ was sent along the $c$ axis of the single crystal (along $z$), so perpendicular to the $CuO_2$ planes, generating a longitudinal temperature difference $\Delta T_z = T+ - T-$ (Extended Data Fig. 1b). By applying a magnetic field $H$ along the $a$ axis of the crystal (along $x$), so parallel to the $CuO_2$ planes, one generates a transverse temperature difference $\Delta T_y$ (Extended Data Fig. 1b). The longitudinal thermal conductivity along the $z$ axis is given by $\kappa_{zz} = (J / \Delta T_z) (L_z / w_y t_x)$, where $L_z$ is the separation (along $z$) between the two points at which $T+$ and $T-$ are measured, $w_y$ is the width of the sample (along $y$) and $t_x$ its thickness (along $x \mathbin{/\mkern-5mu/} a$). The out-of-plane thermal Hall conductivity is defined as $\kappa_{zy} = -\kappa_{yy} (\Delta T_y / \Delta T_z) (L_z / w_y)$, where $\kappa_{yy}$ is the longitudinal thermal conductivity along the $y$ axis (again taken to be equal to $\kappa_{xx}$).

## LONGITUDINAL THERMAL CONDUCTIVITY

As reported previously[4], we find that $La_2CuO_4$ has a nearly isotropic longitudinal thermal conductivity at low temperature (Extended Data Fig. 2). Indeed, $\kappa_{xx} / \kappa_{zz} = \kappa_a / \kappa_c \sim 0.8$ at $T = 25$ K. At higher temperature, thermally excited magnons contribute to heat transport in $La_2CuO_4$, but only within the plane, and $\kappa_a$ grows to exceed $\kappa_c$ (ref. 4). So at low temperature, where phonons dominate the heat transport, the conductivity of phonons is nearly isotropic.

In Nd-LSCO, we also find that the phonon conductivity is nearly isotropic, as can be deduced from the data in Extended Data Fig. 3. Once we remove the contribution of mobile charge carriers, as done in panels 3b and 3d using the Wiedemann-Franz law, we have $\kappa_{xx} / \kappa_{zz} = \kappa_a / \kappa_c \sim 1.2$ and 1.3 at $T = 25$ K, for $p = 0.21$ and $p = 0.24$, respectively.

Note that the electrical conductivity, and therefore also the electronic thermal conductivity from charge carriers, is highly anisotropic, with $\sigma_{xx} / \sigma_{zz} = \rho_{zz} / \rho_{xx} \sim 250$ in Nd-LSCO at $p = 0.24$ (from ref. 33).

## ANISOTROPY OF THE ELECTRONIC HALL CONDUCTIVITY

In Extended Data Fig. 4, we show the anisotropy of the electrical Hall conductivity in Nd-LSCO at $p = 0.24$, plotted as $L_0 \sigma_{ny}$ vs $T$, for $n = x$ (blue) and $n = z$ (red). We see that there is an anisotropy comparable to that found in the longitudinal conductivity, namely $\sigma_{xy} / \sigma_{zy} \sim 250$. We expect that any excitation of electronic origin, arising from either charge or spin degrees of freedom, would yield a similarly strong anisotropy in both its longitudinal and transverse thermal conductivities. For that reason, we attribute the nearly isotropic thermal Hall conductivity found in $La_2CuO_4$ to phonons.

## WIEDEMANN-FRANZ LAW FOR CURRENTS NORMAL TO THE PLANES

In Nd-LSCO with $p = 0.24$, the maximal contribution of charge carriers to $\kappa_{zy}(T)$ can be estimated using the Wiedemann-Franz law, namely $\kappa_{zy} / T \leq L_0 \sigma_{zy}$, where $\sigma_{zy} \leq \rho_{zy} / (\rho_{zz} \rho_{yy})$. The data for $\rho_{zz}$ and $\rho_{zy}$ are displayed in Extended Data Figs. 5a and 5b, respectively. The $\rho_{xx}$ data for the $a$-axis sample cut from the same crystal was reported in ref. 20. The resulting curve for $L_0 \sigma_{zy}$ (using $\rho_{yy} = \rho_{xx}$) is displayed in Extended Data Figs. 5c. Because $\rho_{zz} / \rho_{xx} \sim 250$ in Nd-LSCO at $p = 0.24$, we have a correspondingly very small electrical Hall conductivity $\sigma_{zy}$. The maximal value of the electronic thermal Hall conductivity along the $c$ axis is about $\kappa_{zy} / T \sim 0.01$ mW / K$^2$ m, at $T = 10$ K – a value 200 times smaller than the electronic thermal Hall conductivity measured in the plane (Fig. 1b). In Extended Data Fig. 5d, we see that this maximal electronic $\kappa_{zy}$ is within the noise of our measured $\kappa_{zy}$. Any phonon contribution to $\kappa_{zy}$ in Nd-LSCO at $p = 0.24$ must therefore be smaller than $| \kappa_{zy} / T | \sim 0.01$ mW / K$^2$ m, at $T = 10$ K. This is smaller than the measured $| \kappa_{zy} / T |$ in Nd-LSCO at $p = 0.21$ by a factor $\sim 100$. In other words, the thermal Hall response of phonons in Nd-LSCO undergoes an increase of at least 100-fold immediately upon crossing below $p^*$.

FIELD DEPENDENCE OF THE CONDUCTIVITIES

In Extended Data Fig. 6, we show how $\kappa_{zz}$ and $\kappa_{zy}$ vary with the strength of the applied field, for three of our samples, by plotting : 1) $\kappa_{zz}$ vs $T$ at $H = 0$, 10 and 15 T; 2) $\kappa_{zy} / (T H)$ vs $T$ at $H = 10$ and 15 T. In all cases, the field dependence of the thermal conductivity $\kappa_{zz}$ is very small, and the thermal Hall conductivity $\kappa_{zy}$ is essentially linear in $H$.

BACKGROUND SIGNAL FROM THE SAMPLE MOUNT

Because in our experimental set-up the sample was attached to a block of copper serving as the heat sink (Extended Data Fig. 1), one might expect a thermal Hall signal to come from Cu. However, in all the data reported here and published in ref. 3, all obtained using the same set-up, this background signal from Cu was negligible, as demonstrated in three ways.

First, the fact that the Wiedemann-Franz law is satisfied for in-plane transport in Nd-LSCO $p = 0.24$ (ref. 3) rules out any significant contamination of the $\kappa_{xy}$ data in that measurement.

Secondly, the fact that our $c$-axis data on Nd-LSCO and Eu-LSCO with $p = 0.24$ yield $| \kappa_{zy} / T | < 0.01$ mW / K$^2$ m for all temperatures up to at least 100 K (red, Figs. 1b and 1d) shows that the signal from Cu is smaller than the noise in our measurement.

Thirdly, we have carried out a test study whereby a cuprate sample was measured first in the usual way, using a Cu block to which the sample was glued with Ag paste, and then re-measured using a block of LiF (an insulator known to generate no thermal Hall signal) to which the sample was glued with GE varnish. All other aspects of the experiment were kept the same in the two measurements (contacts, wires, heater, thermometers, electronics). The thermal Hall signal obtained in the two separate ways was identical, within error bars, i.e. $\kappa_{xy}(T)$ was fully reproduced all the way from 10 K to 100 K. This test data will be reported in a separate paper (M.-E. Boulanger *et al.*, in preparation). The smallness of the contamination from the Cu block is largely due to the mounting geometry whereby the main (longitudinal) temperature gradient in the Cu block is perpendicular to the main (longitudinal) temperature gradient in the sample (Extended Data Fig. 1).

PRIOR ARGUMENTS AGAINST A PHONON SCENARIO

In ref. 3, it was considered unlikely that phonons could be responsible for the large negative thermal Hall signal $\kappa_{xy}$ found in cuprates based on two observations. First, the longitudinal phonon thermal conductivity $\kappa_{xx}$ couples very weakly to the magnetic field. Specifically, $\kappa_{xx}$ changes at most by 0.5 % in 15 T, and the maximal ratio $\kappa_{xy} / \kappa_{xx}$ is also about 0.5 %. This is much smaller than in multiferroics[17], for example, where phonons are thought to be responsible for the thermal Hall effect. This weak field dependence of $\kappa_{xx}$ in cuprates is now one aspect of the phenomenology that needs to be understood.

The second observation is that the phonon part of the longitudinal thermal conductivity $\kappa_{xx}$ increases upon crossing below $p^*$ (ref. 3), as opposed to the decrease one might expect if some new scattering mechanism of phonons (causing the chirality) appears in the pseudogap phase. An increase in $\kappa_{xx}$ is natural in view of the large drop in the charge carrier density of Nd-LSCO below $p^*$ (ref. 20), which means that phonons become less scattered by electrons. So a putative extra scattering mechanism would have to overcompensate for the electron-phonon effect. At this stage, the quantitative aspects of these two scattering mechanisms are unknown. Also, it may be that a scattering mechanism is not really what confers chirality to phonons in the pseudogap phase. They may instead acquire a Berry curvature, for example, which may not have a large effect on $\kappa_{xx}$.

CRYSTAL STRUCTURE OF Nd-LSCO

At low temperature, the material $La_{1.6-x}Nd_{0.4}Sr_xCuO_4$ adopts the so-called low-temperature tetragonal (LTT) crystal structure, for a range of Sr concentrations that extends down to at least $x = 0.10$ and up to at least $x = 0.25$ (ref. 34). At $x = 0.20$ and $x = 0.25$, x-ray diffraction detects the structural transition into the LTT phase upon cooling at $T_{LTT} \sim 70$ K and 50 K (ref. 34), respectively (Extended Data Fig. 7a).

At temperatures above $T_{LTT}$, the structural phase is labeled LTO1 (low-temperature orthorhombic), with transitions at $T_{LTO1} \sim 250$ K and 150 K, for $x = 0.20$ and $x = 0.25$, respectively.

So our two Nd-LSCO samples with nominal Sr concentrations $x = 0.21$ and $x = 0.25$, refined to $p = 0.21 \pm 0.003$ and $p = 0.24 \pm 0.005$, are both expected to have the same crystal structure below 150 K. We have confirmed this by performing dilatometry measurements of the sample length $L$ vs temperature $T$ in Nd-LSCO samples with $p = 0.21$ (the actual sample in which $\kappa_{zy}$ was measured) and $p = 0.24$ (a sample cut immediately next to the sample in which $\kappa_{zy}$ was measured). The data are shown in Extended Data Fig. 7b, plotted as $dL / dT$ vs $T$. A clear anomaly is observed in both samples at the structural transition, with transition temperatures $T_{LTT} \sim 82 \pm 5$ K at $p = 0.21$ and $T_{LTT} \sim 45 \pm 10$ K at $p = 0.24$.

This confirms that there is no structural difference between our two Nd-LSCO samples, with $p = 0.21$ and $p = 0.24$. Therefore, this rules out the possibility that the large phonon Hall effect observed at $p = 0.21$, completely absent at $p = 0.24$, is due to structural domain boundaries that scatter phonons, as proposed for $SrTiO_3$ (ref. 18), or to any other structural feature.

**Data availability.** The data that support the plots within this paper and other findings of this study are available from the corresponding author upon reasonable request.

# EXTENDED DATA FIGURE CAPTIONS

**Extended Data Fig. 1 | Current and field orientation for $\kappa_{xy}$ and $\kappa_{zy}$ measurements.**

Sketch of the thermal Hall measurement setup for **a)** $J // a // -x$ and **b)** $J // c // z$. The Cartesian coordinate system is defined in the same way for the two samples.

**Extended Data Fig. 2 | Longitudinal thermal conductivities $\kappa_{xx}$ and $\kappa_{zz}$ in $La_2CuO_4$.**

Thermal conductivity versus temperature in a field of magnitude $H = 15$ T for $La_2CuO_4$ ($p = 0$), plotted as **a)** $\kappa_{nn}$ vs $T$ and **b)** $\kappa_{nn} / T$ vs $T$, for heat current directions $J // a$ ($n = x$; blue) and $J // c$ ($n = z$; red). The longitudinal thermal conductivity of phonons at low temperature is nearly isotropic, with $\kappa_{xx} / \kappa_{zz} = \kappa_a / \kappa_c \sim 0.8$ at $T = 25$ K.

**Extended Data Fig. 3 | Longitudinal thermal conductivities $\kappa_{xx}$ and $\kappa_{zz}$ in Nd-LSCO.**

Thermal conductivity $\kappa_{nn}$ versus temperature in a field of magnitude $H = 15$ T, plotted as $\kappa_{nn} / T$ vs $T$, for **a)** Nd-LSCO with $p = 0.21$ and **b)** Nd-LSCO with $p = 0.24$, for heat current directions $J // a$ ($n = x$; blue) and $J // c$ ($n = z$; red). In panels c) and d), the thermal conductivity of charge carriers is subtracted, using the Wiedemann-Franz law to estimate its magnitude (see ref. 3). The longitudinal thermal conductivity of phonons at low temperature is nearly isotropic, with $\kappa_{xx} / \kappa_{zz} = \kappa_a / \kappa_c \sim 1.2$ and 1.3 at $T = 25$ K, for $p = 0.21$ and $p = 0.24$, respectively.

**Extended Data Fig. 4 | Anisotropy of electrical Hall conductivity in Nd-LSCO.**

Electrical Hall conductivity $\sigma_{ny}$ versus temperature in a field of magnitude $H = 15$ T, plotted as $L_0\, \sigma_{ny}$ vs $T$, for Nd-LSCO with $p = 0.24$, for heat current directions $J // a$ ($n = x$; blue) and $J // c$ ($n = z$; red). The data for $\sigma_{zy}$ are multiplied by a factor 250. We use the approximate relation $\sigma_{zy} = \rho_{zy} / (\rho_{zz}\, \rho_{xx})$, with $\rho_{zz}$ and $\rho_{xx}$ data taken from ref. 33. The $\sigma_{xy}$ data are taken from ref. 3.

**Extended Data Fig. 5 | Electronic thermal Hall conductivity in Nd-LSCO $p = 0.24$.**

Estimate of the maximal $c$-axis thermal Hall conductivity from charge carriers in our sample of Nd-LSCO with $p = 0.24$, obtained by applying the Wiedemann-Franz law to the measured electrical Hall conductivity $\sigma_{zy}$, namely $\kappa_{zy} / T \leq L_0\, \sigma_{zy}$, where $\sigma_{zy} \leq \rho_{zy} / (\rho_{zz}\, \rho_{yy})$. **a)** Electrical resistivity for $J // c$, $\rho_{zz}$ vs $T$ (from ref. 33); **b)** electrical Hall resistivity for $J \,//\, c$ and $H \,//\, a$, $\rho_{zy}$ vs $T$ ; **c)** maximal electrical Hall conductivity for $J \,//\, c$ and $H \,//\, a$, defined as $\sigma_{zy} = \rho_{zy} / (\rho_{zz} \rho_{xx})$ (with $\rho_{xx}$ data from ref. 33), plotted as $L_0\, \sigma_{zy}$ (multiplied by 60) vs $T$ ; **d)** comparison of the measured electrical ($\sigma_{zy}$) and thermal ($\kappa_{zy}$) Hall conductivities, plotted as $L_0\, \sigma_{zy}$ (blue) and $\kappa_{zy} / T$ (red; Fig. 1b) vs $T$.

**Extended Data Fig. 6 | Field dependence of $\kappa_{zz}$ and $\kappa_{zy}$ in La$_2$CuO$_4$ and Nd-LSCO.**

*Upper panels*: thermal conductivity $\kappa_{zz}$ measured at $H = 0$ T (purple), 10 T (green) and 15 T (red), plotted as $\kappa_{zz} / T$ vs $T$, for **a)** La$_2$CuO$_4$, **b)** Nd-LSCO with $p = 0.21$ and **c)** Nd-LSCO with $p = 0.24$. *Lower panels*: thermal Hall conductivity $\kappa_{zy}$ measured at $H = 10$ T (green) and 15 T (red), plotted as $\kappa_{zy} / (T\, H)$ vs $T$, for **d)** La$_2$CuO$_4$, **e)** Nd-LSCO with $p = 0.21$ and **f)** Nd-LSCO with $p = 0.24$. We see that $\kappa_{zy}$ is approximately linear in $H$.

**Extended Data Fig. 7 | Structural transition in Nd-LSCO.**

**a)** Structural phase diagram of Nd-LSCO as a function of doping. The black dots and black line mark the structural transition from the LTO1 phase to the LTT phase at low temperature, at $T_{LTT}$, as measured by x-ray diffraction[34]. The squares mark $T_{LTT}$ in our samples with $p = 0.21$ (blue) and $p = 0.24$ (green), as detected by dilatometry measurements (see panel b). **b)** Change in sample length $L$ as a function of temperature, plotted as its derivative $dL / dT$ vs $T$, measured in our $c$-axis sample of Nd-LSCO with $p = 0.21$ (blue) and in a sample of Nd-LSCO cut from the same large single crystal as, and next to, our $c$-axis sample of Nd-LSCO with $p = 0.24$ (green). The dip in the curves marks the structural phase transition from the LTO1 phase above to the LTT phase below the transition temperature $T_{LTT}$, where $T_{LTT} = 82 \pm 5$ K at $p = 0.21$ and $T_{LTT} = 45 \pm 10$ K at $p = 0.24$. These data confirm that our two Nd-LSCO samples, with $p = 0.21$ and $p = 0.24$, have the same crystal structure. This shows that all the differences observed in their properties, in particular the dramatic difference in their thermal Hall conductivity $\kappa_{zy}$ (Fig. 1), are not due to a difference in structural properties. Instead, these differences are linked with the onset of the pseudogap phase at $p^* = 0.23$.

[1] Keimer, B. *et al.* From quantum matter to high-temperature superconductivity in copper oxides. *Nature* **518**, 179-186 (2015).

[2] Proust, C. & Taillefer, L. The remarkable underlying ground states of cuprate superconductors. *Annu. Rev. Condens. Matter Phys.* **10**, 409-429 (2019).

[3] Grissonnanche, G. *et al.* Giant thermal Hall conductivity in the pseudogap phase of cuprate superconductors. *Nature* **571**, 376-380 (2019).

[4] Hess, C. *et al.* Magnon heat transport in doped $La_2CuO_4$. *Phys. Rev. Lett.* **90**, 197002 (2003).

[5] Ye, M. *et al*. Phonon dynamics in the Kitaev spin liquid. arXiv:2002.05328 (2020).

[6] Hentrich, R. *et al.* Large thermal Hall effect in α-$RuCl_3$ : Evidence for heat transport by Kitaev-Heisenberg paramagnons. *Phys. Rev. B* **99**, 085136 (2019).

[7] Hirschberger, M. *et al*. Large thermal Hall conductivity of neutral spin excitations in a frustrated quantum magnet. *Science* **348**, 106–109 (2015).

[8] Onose, M. *et al.* Observation of the magnon Hall effect. *Science* **329**, 297 (2010).

[9] Katsura, H., Nagaosa, N. & Lee, P. A. Theory of the thermal Hall effect in quantum magnets. *Phys. Rev. Lett.* **104**, 066403 (2010).

[10] Lee, H., Han, J. H. & Lee, P. A. Thermal Hall effect of spins in a paramagnet. *Phys. Rev. B* **91**, 125413 (2015).

[11] Qin, T., Zhou, J. & Shi, J. Berry curvature and the phonon Hall effect. *Phys. Rev. B* **86**, 104305 (2012).

[12] Yang, Y.-F., Zhang, G.-M, & Zhang, F.-C. Universal behavior of the anomalous thermal Hall conductivity. arXiv:2001.08583 (2020).

[13] Nasu, J., Yoshitake, J. & Motome, Y. Thermal transport in the Kitaev model. *Phys. Rev. Lett.* **119**, 127204 (2017).

[14] Ye, M. *et al.* Quantization of the thermal Hall conductivity at small Hall angles. *Phys. Rev. Lett.* **121**, 147201 (2018).

[15] Vinkler-Aviv, Y. & Rosch, A. Approximately quantized thermal Hall effect of chiral liquids coupled to phonons. *Phys. Rev. X* **8**, 031032 (2018).

[16] Mori, M. *et al.* Origin of the phonon Hall effect in rare-earth garnets. *Phys. Rev. Lett.* **113**, 265901 (2014).

[17] Ideue, T. *et al.* Giant thermal Hall effect in multiferroics. *Nature Mat.* **16**, 797-802 (2017).

[18] Li, X. *et al*. Phonon thermal Hall effect in strontium titanate. arXiv:1909.06552 (2019).

[19] Samajdar, R. *et al.* Enhanced thermal Hall effect in the square-lattice Néel state. *Nat. Phys.* **15**, 1290-1294 (2019).

[20] Collignon, C. *et al.* Fermi-surface transformation across the pseudogap critical point of the cuprate superconductor $La_{2-x}Nd_{0.4}Sr_xCuO_4$. *Phys. Rev. B* **95**, 224517 (2017).

[21] Matt, C. E. *et al*. Electron scattering, charge order and pseudogap physics in $La_{2-x}Nd_{0.4}Sr_xCuO_4$ : an angle-resolved photoemission spectroscopy study. *Phys. Rev. B* **92**, 134524 (2015).

[22] Michon, B. *et al.* Thermodynamic signatures of quantum criticality in cuprate superconductors. *Nature* **567**, 281-222 (2019).

[23] Strohm, C., Rikken, G. L. J. A. & Wyder, P. Phenomenological evidence for the phonon Hall effect. *Phys. Rev. Lett.* **95**, 155901 (2005).

[24] He, Y. *et al.* Rapid change in superconductivity and electron-phonon coupling through the critical point in Bi-2212. *Science* **362**, 62-65 (2019).

[25] Chen, J.-Y., Kivelson, S. A., & Sun, X.-Q. Enhanced thermal Hall effect in nearly ferroelectric insulators. arXiv:1910.00018 (2019).

[26] Kyung, B. *et al.* Pseudogap induced by short-range spin correlations in a doped Mott insulator. *Phys. Rev. B* **73**, 165114 (2006).

[27] Scheurer, M. S. *et al.* Topological order in the pseudogap metal. *PNAS* **115**, E3665-E3672 (2018).

[28] Hunt, A. W. *et al.* Glassy slowing of stripe modulation in $(La,Eu,Nd)_{2-x}(Sr,Ba)_xCuO_4$: a $^{63}Cu$ and $^{139}La$ NQR study down to 350 mK. *Phys. Rev. B* **64**, 134525 (2001).

[29] Frachet, M. *et al.* Hidden magnetism at the pseudogap critical point of a high-temperature superconductor. arXiv:1909.10258 (2019).

[30] Kasahara, Y. *et al.* Majorana quantization and half-integer thermal quantum Hall effect in a Kitaev spin liquid. *Nature* **559**, 227-231 (2018).

[31] Cyr-Choinière, O. *et al.* Pseudogap temperature $T^*$ of cuprate superconductors from the Nernst effect. *Phys. Rev. B* **97**, 064502 (2018).

[32] Grissonnanche, G. *et al.* Wiedemann-Franz law in the underdoped cuprate superconductor YBCO. *Phys. Rev. B* **93**, 064513 (2016).

[33] Daou, R. *et al.* Linear temperature dependence of resistivity and change in the Fermi surface at the pseudogap critical point of a high-$T_c$ superconductor. *Nat. Phys.* **5**, 31-34 (2009).

[34] Axe, J. D. & Crawford, M. K. Structural instabilities in lanthanum cuprate superconductors. *J. Low Temp. Phys.* **95**, 271-284 (1994).

**Fig. 1 | Thermal Hall conductivity of cuprates at three different dopings.**

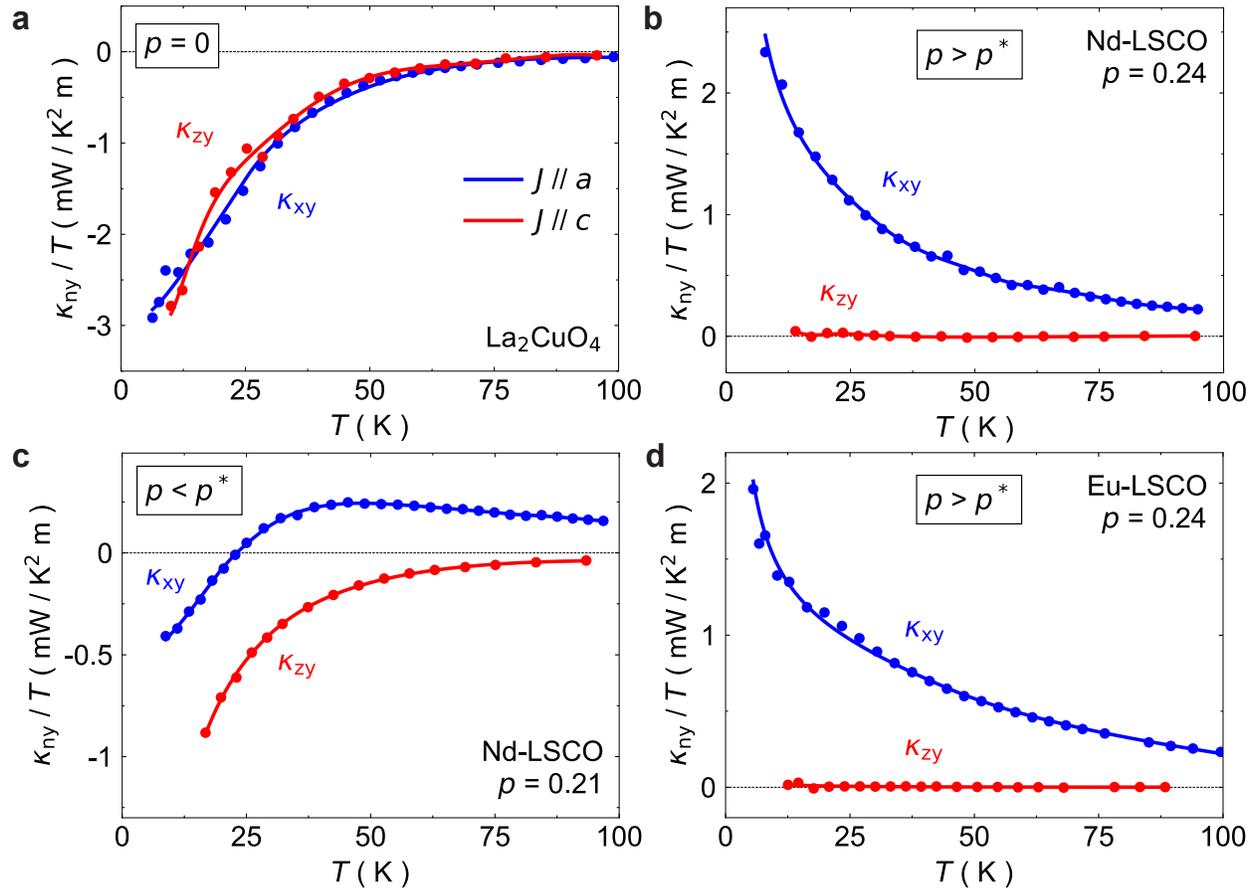

**Fig. 2 | Evolution of the *c*-axis thermal Hall conductivity across the phase diagram.**

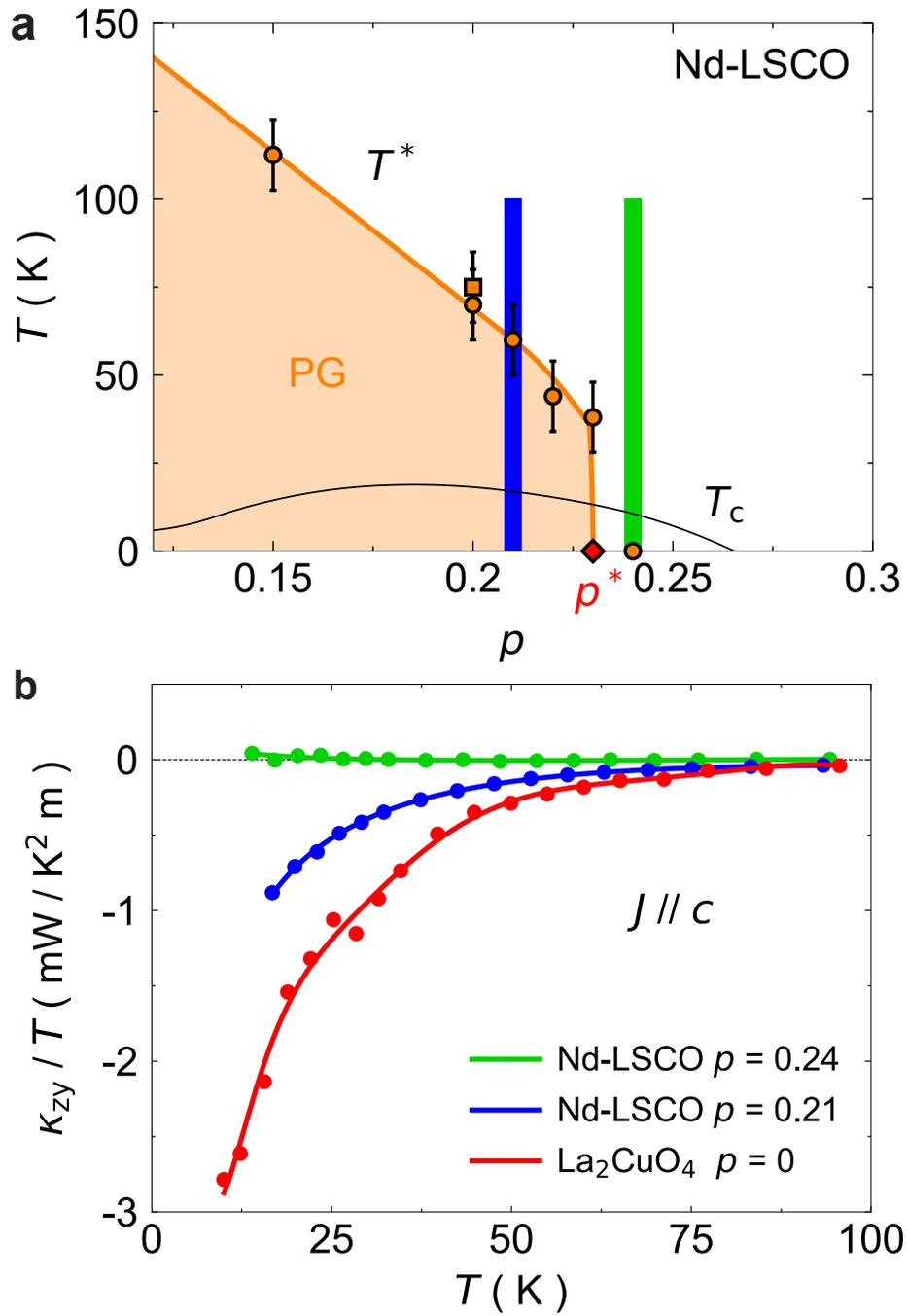

**Fig. 3 | Phenomenological fit to the phonon thermal Hall conductivity.**

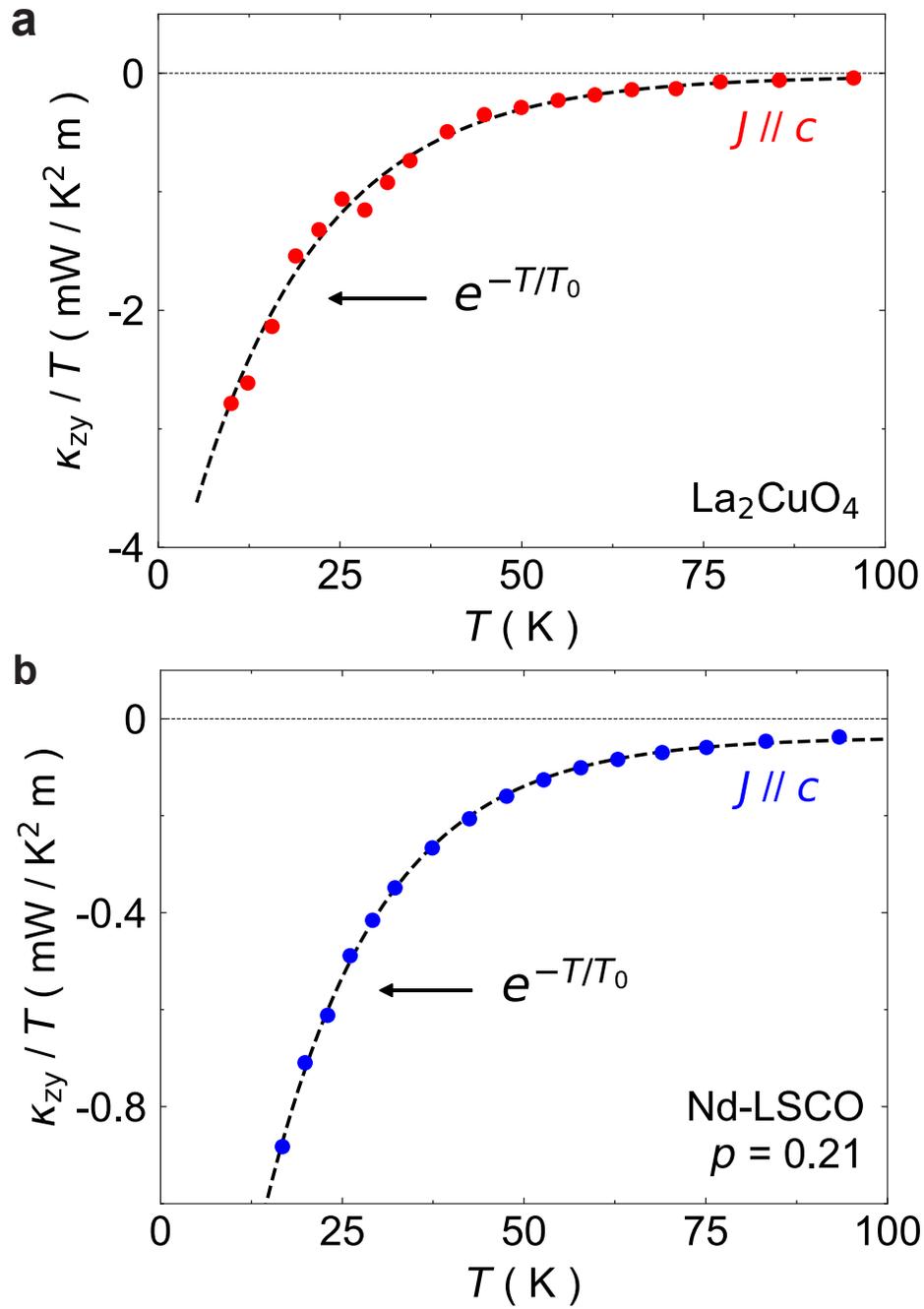

# EXTENDED DATA FIGURES

**Extended Data Fig. 1 | Current and field orientation for $\kappa_{xy}$ and $\kappa_{zy}$ measurements.**

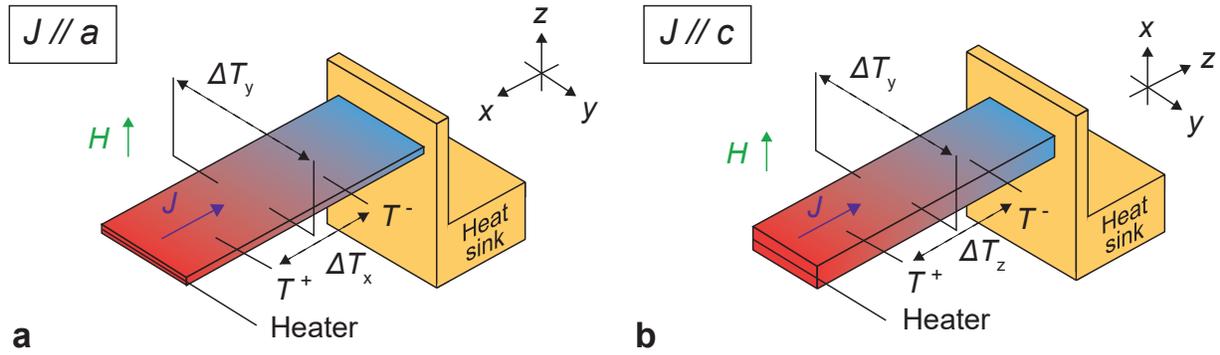

**Extended Data Fig. 2 | Longitudinal thermal conductivities $\kappa_{xx}$ and $\kappa_{zz}$ in La$_2$CuO$_4$.**

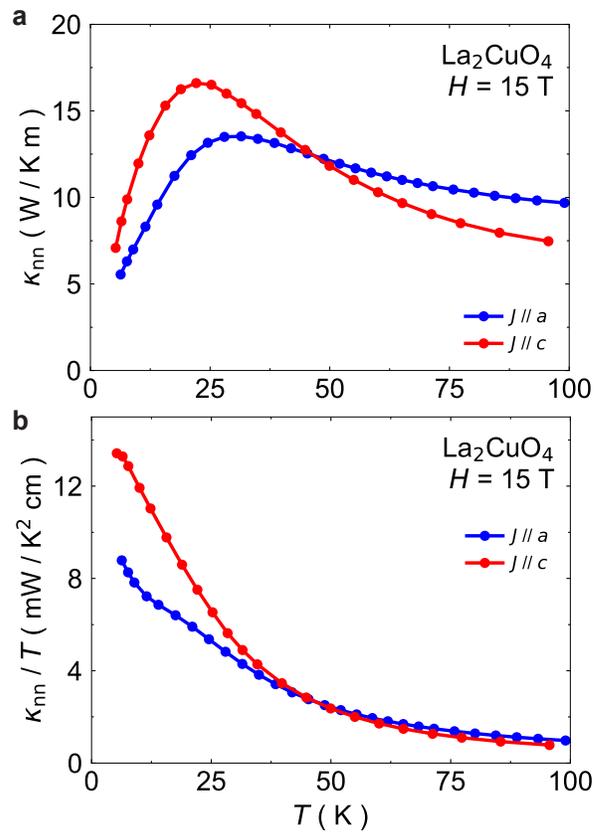

**Extended Data Fig. 3 | Longitudinal thermal conductivities $\kappa_{xx}$ and $\kappa_{zz}$ in Nd-LSCO.**

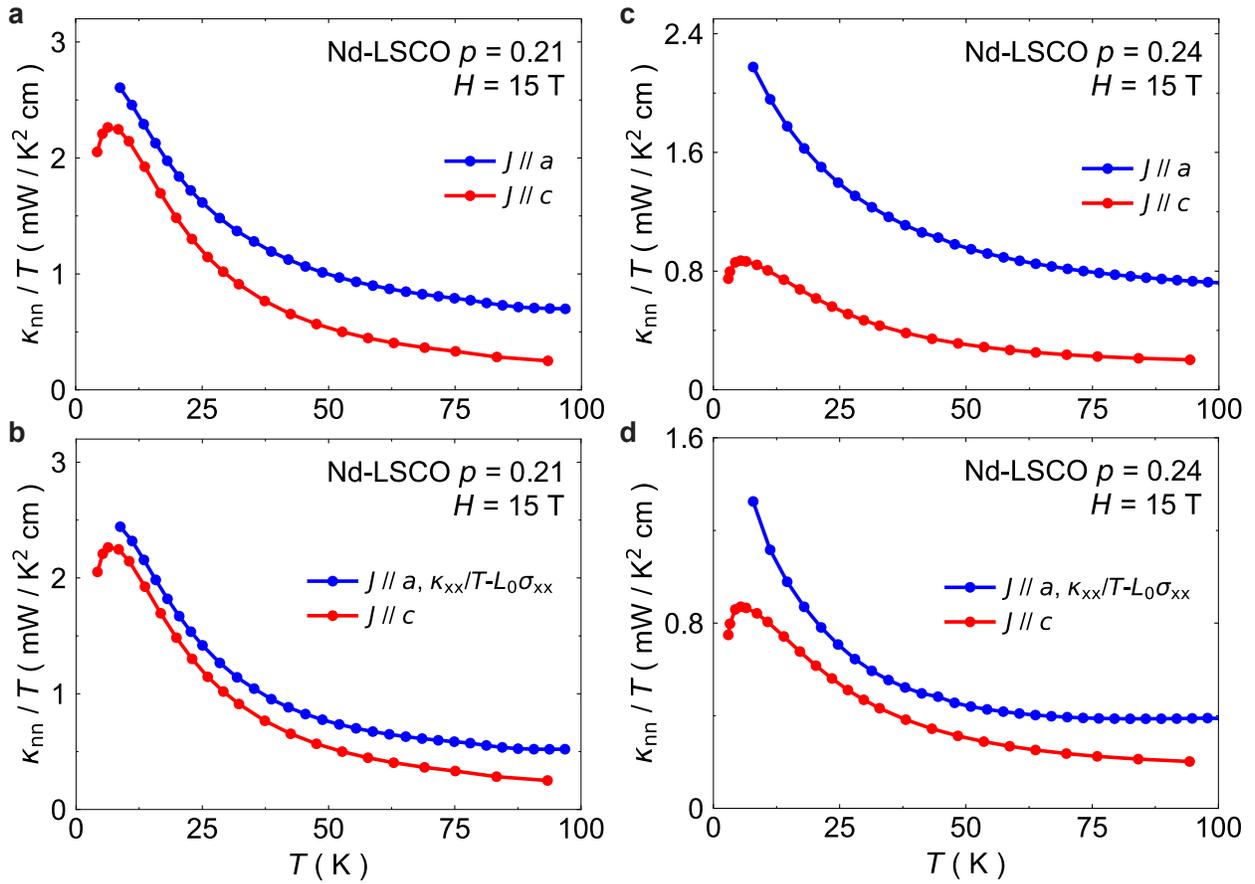

**Extended Data Fig. 4 | Anisotropy of electrical Hall conductivity in Nd-LSCO.**

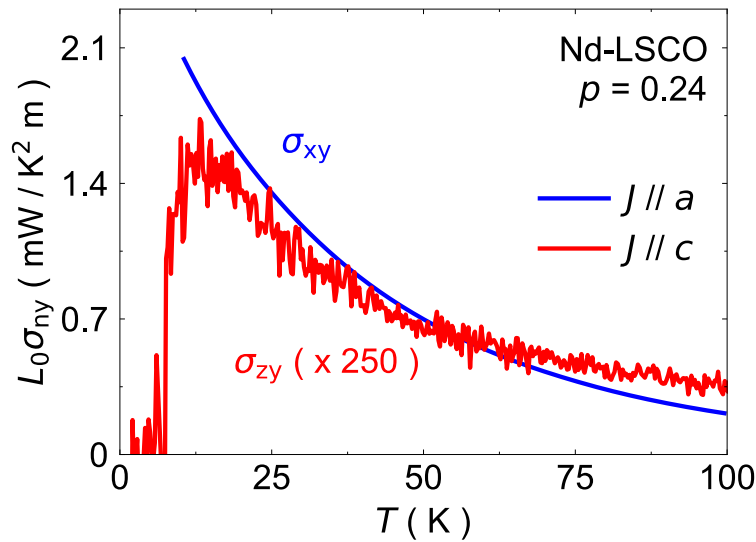

**Extended Data Fig. 5 | Electronic thermal Hall conductivity in Nd-LSCO $p = 0.24$.**

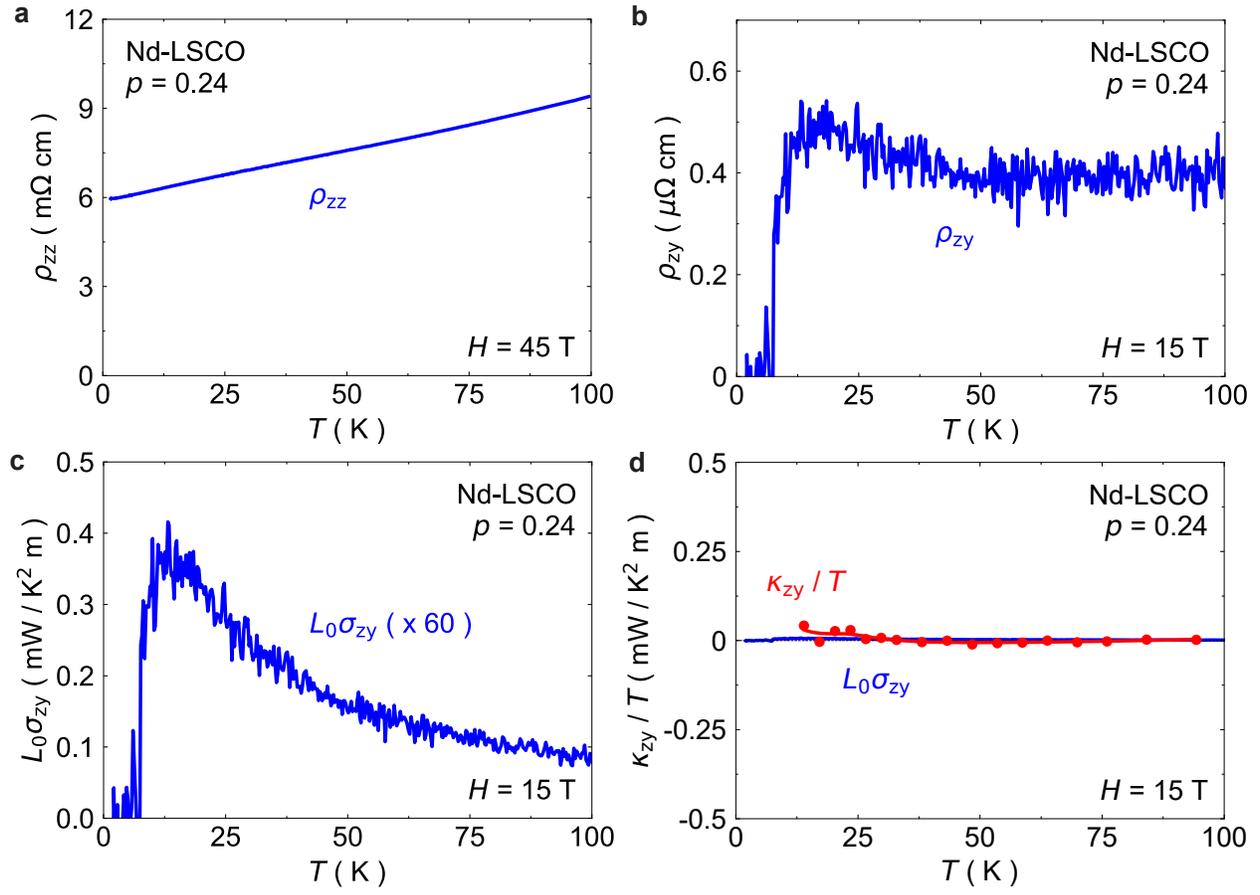

**Extended Data Fig. 6 | Field dependence of $\kappa_{zz}$ and $\kappa_{zy}$ in La$_2$CuO$_4$ and Nd-LSCO.**

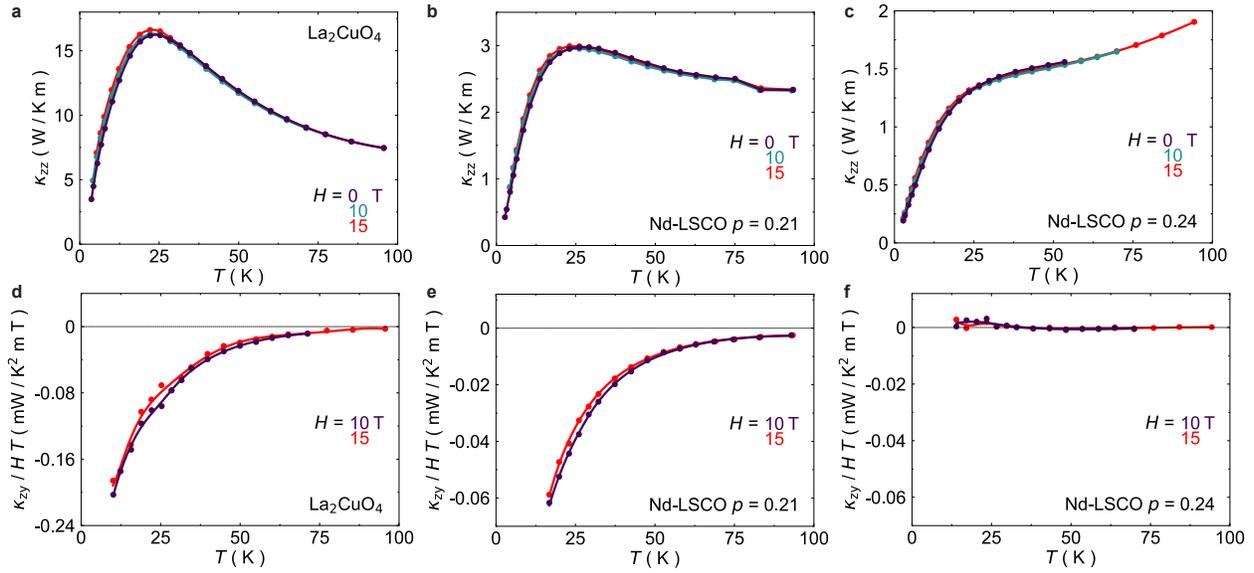

**Extended Data Fig. 7 | Structural transition in Nd-LSCO.**

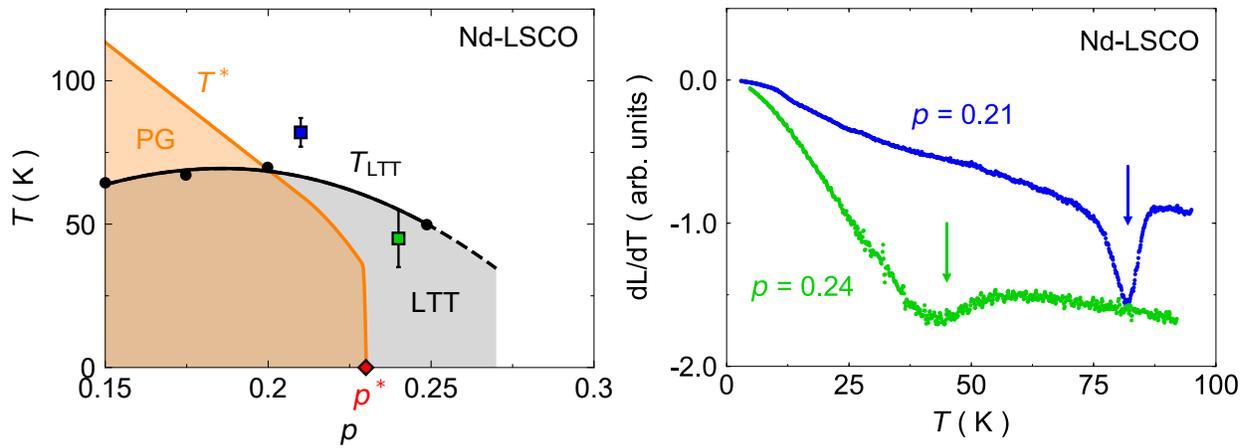